\documentclass[preprint,12pt]{elsarticle}
\usepackage[utf8]{inputenc}
\usepackage{amsmath}
\usepackage{amsmath,amsthm}
\usepackage{amssymb}
\usepackage{genyoungtabtikz}
\usepackage[utf8]{inputenc}
\usepackage[english]{babel}
\usepackage{enumitem}
\usepackage{mathtools}
\usepackage{verbatim}
\usepackage{parskip}
\usepackage{float}
\usepackage{tikz}
\usetikzlibrary{decorations.markings}
\usepackage{adjustbox}
\usepackage{ytableau}
\usepackage{xcolor}

\usepackage[T1]{fontenc}

\usetikzlibrary{matrix, calc, arrows}
\usepackage{comment}
\usepackage[bookmarksopen=false,pdftex=true,breaklinks=true,%
      backref=page,pagebackref=true,plainpages=false,%
      hyperindex=true,pdfstartview=FitH,colorlinks=true,%
      pdfpagelabels=true,colorlinks=true,linkcolor=blue,%
      citecolor=red,urlcolor=green,hypertexnames=false%
      ]%
   {hyperref}

\Yboxdim{13pt}
\Ylinethick{0.6pt}
\Yfillcolor{gray!10}
\Ylinecolor{black!50}

\usepackage{graphicx}
\DeclareMathOperator{\Imm}{Imm}

\newtheorem{theorem}{Theorem}[section]

\newtheorem{lemma}[theorem]{Lemma}
\newtheorem{definition}{Definition}

\newtheorem{example}[theorem]{Example}
\newtheorem{const}{Construction}
\newtheorem{remark}{Remark}
\newtheorem{proposition}[theorem]{Proposition}
\usepackage[margin=1.1in]{geometry}
\usepackage{setspace}
\begin{document}

%\maketitle

\begin{frontmatter}

\title{\#P-hardness proofs of matrix immanants evaluated on restricted matrices}

\author[inst1,inst2]{Istvan Miklos\corref{cor1}}
\ead{miklos.istvan@renyi.hun-ren.hu}
\cortext[cor1]{Corresponding author}

\author[inst3]{Cordian Riener}

\address[inst1]{HUN-REN R\'enyi Institute, 1053 Budapest, Re\'altanoda u. 13-15, Hungary}
\address[inst2]{HUN-REN SZTAKI, 1111 Budapest, L\'agym\'anyosi u. 11, Hungary}
\address[inst3]{Dept. of Mathematics and Statistics, UiT The Arctic University of Norway, Troms\o, Norway}
\begin{abstract}
We establish the $\#P$-hardness of computing a broad class of immanants, even when restricted to specific categories of matrices. Concretely, we prove that computing $\lambda$-immanants of $0$-$1$ matrices is $\#P$-hard whenever the partition~$\lambda$ contains a sufficiently large domino-tileable region, subject to certain technical conditions.

We also give hardness proofs for some $\lambda$-immanants of weighted adjacency
matrices of planar directed graphs, such that the shape $\lambda =
(\mathbf{1} + \lambda_d)$ has size $n$ such that $|\lambda_d| = n^\varepsilon$
for some $0 < \varepsilon < \frac{1}{2}$, and such that for some $w$, the shape $\lambda_d/(w)$ is tileable with $1 \times 2$ dominos.
\end{abstract}


\end{frontmatter}

%\begin{abstract}
%We establish the $\#P$-hardness of computing a broad class of immanants, even when restricted to specific categories of matrices. Concretely, we prove that computing $\lambda$-immanants of $0$-$1$ matrices is $\#P$-hard whenever the partition~$\lambda$ contains a sufficiently large domino-tileable region, subject to certain technical conditions. 

%We also give hardness proofs for some $\lambda$-immanants of weighted adjacency
%matrices of planar directed graphs, such that the shape $\lambda =
%(\mathbf{1} + \lambda_d)$ has size $n$ such that $|\lambda_d| = n^\varepsilon$
%for some $0 < \varepsilon < \frac{1}{2}$, and such that for some $w$, the shape $\lambda_d/(w)$ is tileable with $1 \times 2$ dominos.

%\end{abstract}

% \begin{keyword}
%\#P-hardness \sep immanants \sep computational complexity \sep matrix theory
%\end{keyword}
 
\section{Introduction}
A matrix functional is a mapping that assigns a scalar value to an $n\times n$ matrix $M := \{m_{i,j}\}$ with entries in a commutative ring of characteristic zero.  
%Remark (Istvan): Our theorems here holds for rings with 0 characteristic. I like the idea to define matrix functions for any commutative ring but it might cause more complications than what we would otherwise gain...
%Yes, I (cordi) agree. I was just wuondering. Sometimes there is no harm in doing things more generally. But actually I think here the gain is very limited and in general if some other people will need this, they can easily generalize this 
Two well-known examples of matrix functionals are the \emph{determinant}
\[
\det(M) := \sum_{\rho \in S_n} \mathrm{sign}(\rho)\, \prod_{i=1}^n m_{i,\rho(i)}
\]
and the \emph{permanent}
\[
\mathrm{per}(M) := \sum_{\rho \in S_n} \prod_{i=1}^n m_{i,\rho(i)},
\]
where \(S_n\) denotes the symmetric group on \(n\) elements. Despite their similar definitions,
 %-both being sums over \(n!\) terms -
 the two functionals behave very differently. In particular, they exhibit strikingly different computational complexities: while the determinant of an \(n \times n\) matrix can be computed in \(O(n^{2.373})\) arithmetic operations~\cite{LeGall2014,JW2020}, computing the permanent is \#P-hard~\cite{Valiant-permanent}.

The contrast between these two functionals was observed well before Valiant’s seminal paper. In 1913, Pólya~\cite{Polya} asked whether the permanent could be obtained from the determinant by uniformly multiplying the entries of the matrix by $+1$ and $-1$. Szegő~\cite{szego} subsequently showed that such a conversion is impossible for matrices larger than \(2 \times 2\).

Valiant conjectured that computing the permanent requires a super-polynomial number of arithmetic operations over any field of characteristic different from~$2$. B\"urgisser proved that if Valiant's conjecture fails over some field of positive characteristic, or if it fails over a field of characteristic zero under the assumption of the Generalized Riemann Hypothesis, then $\mathrm{NP} \subseteq \mathrm{P}/\mathrm{poly}$~\cite{burgisser-on-valiant}.

%In both the determinant and the permanent,
In both the definition of the determinant and the definition of the permanent,
the symmetric group plays a central role: the involved terms, as well as their coefficients, are defined via permutations in~$S_n$. Building on this observation, Littlewood and Richardson introduced \emph{matrix immanants} as a common generalization of the determinant and the permanent~\cite{LR1934}, using ideas from representation theory. Indeed, the sign character and the trivial (constant $1$) character are both irreducible characters of $S_n$. It is therefore natural to define analogous matrix functionals for an arbitrary irreducible character of $S_n$, leading to the family of functionals known as \emph{matrix immanants}. The determinant and the permanent arise as the special cases corresponding to the sign and trivial characters, respectively.

Before discussing the main results of this paper, we give a brief overview of irreducible characters of the symmetric group and recall the formal definition of immanants. For completeness, we also recall the definition of permutations and their cycle structure.

%Maybe it is good to structure with subsections. In this way readers familiar with all this can easily skipe it
\subsection{Irreducible $S_n$ representations,  characters, and immanants}
%%%%%%%%%%%%%%%%%%%%%%%
In this section we introduce the key players of this article: 
the symmetric group $S_n$, its irreducible representations and their characters, 
and finally, the associated matrix functions, called \emph{immanants}. 
We begin with basic notions on permutations before turning to representation theory and the link between characters and immanants.

\begin{definition}
Let $n\in\mathbb{N}$ be a natural number. 
\begin{enumerate}
    \item A \emph{permutation} of length $n$ is a bijection $\sigma: [n] = \{1,2,\dots,n\} \to [n]$.  
    \item Every permutation can be uniquely written (up to the ordering of factors) as a product of disjoint \emph{cycles}. A cycle of length $k$, denoted  
    \[
    (c_1\, c_2\, \ldots\, c_k),
    \]
    means $\sigma(c_i) = c_{i+1}$ for $1 \le i < k$ and $\sigma(c_k) = c_1$.  
    \item The \emph{cycle structure} of $\sigma$ is the multiset of lengths of its disjoint cycles.  
    \item The set of all the $n!$ permutations of $[n]$, with composition as the group operation, forms the \emph{symmetric group} $S_n$.  
\end{enumerate}
\end{definition}

\medskip

We next recall some basic notions from representation theory (see \cite{fulton,JamesKerber} for details).

\begin{definition}
Let $G$ be a finite group. 
\begin{enumerate}
    \item A (complex) \emph{representation} of $G$ is a homomorphism
\[
\varphi:\, G \longrightarrow \mathrm{GL}(V),
\]
where $V$ is a complex vector space, i.e., a map embedding $G$ into the group of invertible linear transformations of $V$.  

\item The associated \emph{character} is the map
\[
\chi_\varphi: G \longrightarrow \mathbb{C}, \qquad g \mapsto \operatorname{Tr}(\varphi(g)),
\]
where $\operatorname{Tr}$ denotes the trace.  

\item A representation is called \emph{irreducible} if $V$ contains no non-trivial $G$-invariant subspace. The characters of irreducible representations are called \emph{irreducible characters}.
\end{enumerate}
\end{definition}
\begin{remark}
Since $\operatorname{Tr}(AB)=\operatorname{Tr}(BA)$ for square matrices $A,B$, it follows that
\[
\operatorname{Tr}(\varphi(g)) = \operatorname{Tr}(\varphi(h^{-1}hg)) = \operatorname{Tr}(\varphi(h^{-1})\varphi(h)\varphi(g))  
= \operatorname{Tr}\!\bigl(\varphi(h)\varphi(g)\varphi(h^{-1})\bigr) 
= \operatorname{Tr}(\varphi(hgh^{-1}))
\]
for any $g,h\in G$. Thus, characters are independent of the chosen basis and are constant on conjugacy classes of $G$, i.e., they are \emph{class functions}. Moreover, the irreducible characters form a basis of the vector space of all class functions. In particular, the number of irreducible characters equals the number of conjugacy classes.
\end{remark}
\begin{example}
The following two irreducible characters exist for every symmetric group $S_n$
\begin{itemize}
    \item The \emph{trivial character}, which assigns $1$ to every permutation.
    \item The \emph{sign character}, which assigns to $\rho \in S_n$ the value $(-1)^{n-c(\rho)}$, where $c(\rho)$ is the number of cycles in the cycle decomposition of $\rho$. Equivalently, this character records the parity of the permutation.
\end{itemize}
\end{example}

\medskip

With these notions, one can define \emph{matrix immanants}, which generalize both the determinant and the permanent using irreducible characters of $S_n$.

\begin{definition}
Let $\varphi$ be an irreducible representation of the symmetric group $S_n$, and let $A=(a_{i,j})$ be an $n\times n$ complex matrix. The \emph{$\varphi$-immanant} of $A$ is
\[
\operatorname{\Imm}_{\varphi}(A) := \sum_{\rho\in S_n} \chi_\varphi(\rho)\, \prod_{i=1}^n a_{i,\rho(i)}.
\]
\end{definition}

It is easy to see that two permutations are in the same conjugacy classes of $S_n$ if and only if they have the same cycle structure. Since characters are constant on conjugacy classes, and since we are mainly interested in their values,  we will  slightly abuse notation using $\rho$ both for the permutation and its  cycle structure. From a combinatorial viewpoint, every cycle structure can be identified with a partition of $n$. Further, the number of conjugacy classes of $S_n$ is therefore  $p(n)$, the number of partitions of the natural number $n$. Since the number of irreducible representations equals the number of the conjugacy classes, $S_n$ has $p(n)$ irreducible representations. To describe partitions in a convenient way we will use the following definitions  which we recall for the convenience of the reader. 

\begin{definition}
A weakly decreasing sequence of integers $\lambda= (\lambda_1, \lambda_2,\ldots, \lambda_h)$ is a \emph{partition} of $n$ if $\sum_{i=1}^h \lambda_i = n$. The \emph{height} of the partition is $h$ and it is denoted by $h(\lambda)$. The size of $\lambda$ is $n$, and it is denoted by $|\lambda|$. Finally, we will denote $n-h$ by $b(\lambda)$.

For short, we will write $c^k$ for a run of $k$ many of $c$'s in a partition. For example, the partition $(2,2,1,\ldots,1)$ of size $n$ can be written as $(2^2,1^{n-4})$. For a series of sequences (even possibly with different lengths), $\lambda$ and $\mu$, we denote by $\lambda+\mu$ the sequence that contains $\lambda_i+\mu_i$ in each position where both $\lambda_i$ and $\mu_i$ exist and $\lambda_i$ or $\mu_i$ where only one of the numbers exists. We will use the addition operation in extending a partition. That is, for partitions $\lambda = (\lambda_1, \lambda_2,\ldots)$ and $\mu = (\mu_1, \mu_2,\ldots)$ we define $(w,\lambda+\mu)$ as $(w,\lambda_1+\mu_1, \lambda_2+\mu_2,\ldots)$.
\end{definition}

\begin{definition}
A partition $\lambda= (\lambda_1, \lambda_2,\ldots, \lambda_h)$ can be represented by a \emph{Young diagram}, which is a finite collection of boxes, arranged into left-justified rows, such that for each $i$, row $i$ contains $\lambda_i$ boxes. A Young diagram is also called a \emph{shape}. 

If $\lambda$ is a shape then its \emph{conjugate} shape $\lambda^*$ is obtained by swapping the rows and columns.
\end{definition}

\begin{example}
Let $n=4$ and $\lambda=(2,1,1)$. A Young diagram of shape $\lambda$ and its conjugate are
\[
  {\yng(2,1,1)}
  \xrightarrow{\ \text{transpose}\ }
  {\yng(3,1)}.
\]
Hence $\lambda^\ast=(3,1)$.
\end{example}

We will also be needing the notion of a skew-shape associated to a pair of partitions.
\begin{definition}
If $\lambda$ and $\mu$ are two partitions, such that $h(\lambda)\ge h(\mu)$, and furthermore, for each $i$ smaller or equal the height of $\mu$, it holds that $\mu_i \le\lambda_i$, then we can define the \emph{skew-shape} $\lambda/\mu$, which is the set-theoretic difference of the Young diagrams of $\lambda$ and $\mu$. The height of a skew-shape $\lambda/\mu$ is the height of $\lambda$. A skew-shape is \emph{connected} if there is a rook path (that is, a path consisting of only horizontal and vertical steps) between any pair of boxes, that is, from any box any other one is reachable by horizontal and vertical steps.
\end{definition}
\begin{example}
Consider the partitions $\lambda=(4,3,1)$ and $\mu=(1,1)$. Then the skew shape $\lambda/\mu$ is 
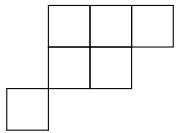
\[
  {
   \gyoung(:;;;,:;;,;)}
\]
Note that this skew-shape is not rook-path connected, that is, it is not a connected skew-shape by definition.
\end{example}

Following the work of Frobenius \cite{Frobenius1900}, the irreducible characters of the symmetric group can be completely understood using these Young diagrams and thus one has a precise way to identify these characters with the partitions, respectively with the Young diagrams. 
For example, the trivial character of $S_n$ is identified with the partition $(n)$ or equivalently with the Young diagram consisting of one long row and the sign character with the partition  $(1^n)$, or equivalent the Young diagram consisting of one long column. Therefore, we can call the immanant constructed  with the irreducible character corresponding to the partition $\lambda$ the $\lambda$-immanant. In this way, the determinant is the $(1^n)$-immanant, and the permanent is the $(n)$-immanant.

Finally a classical result from combinatorics uses Young diagrams to define a partial ordering on the partitions. 
\begin{definition} Two  partitions $\lambda$ and $\mu$  of $n$ are comparable, and $\lambda \preceq \mu$ if $\lambda = \mu$ or if the Young diagram of $\mu$ can be obtained by replacing one or more of the boxes of the Young diagram of $\lambda$ such a way so that each such box can only be moved to the right and upwards.
\end{definition}

According to this partial ordering, we have that $(n)$ is minimal and $(1^n)$ is
maximal.
From this point of view, the determinant and the permanent are maximally distant. Indeed, the determinant corresponds to partitioning $n$ into $n$ parts, while the permanent corresponds to partitioning $n$ into one part. It is therefore  natural to ask how the computational complexity of the immanants changes from the partition $(1^n)$ to the partition $(n)$.

Since we consider some restricted matrices defined as adjacency matrices of graphs, it is important to define certain graphs and their adjacency matrices. 
\begin{definition}
    A \emph{bipartite graph} $G =(U,V,E)$ is a simple graph in which each edge is incident to exactly one vertex from each of the vertex classes $U$ and $V$. The \emph{adjacency matrix} $A$ of a bipartite graph $G =(U,V,E)$ is a $|U|\times|V|$ matrix, in which $a_{i,j} = 1$ if $(u_i,v_j)\in E$ and otherwise $0$. If the bipartite graph $G = (U,V,E)$ is an edge-weighted graph with a weight function $w: E \to \mathbb{C}$, then in the weighted adjacency matrix of $G$, it holds that $a_{i,j}=w(e)$ for each edge $e = (u_i,v_j)$. A bipartite graph is \emph{planar} if it has a drawing in the Euclidean plane without crossing edges.

    A \emph{directed graph} $\vec{G} = (V,E)$ is a graph in which each vertex has a direction, that is, each edge has a head and a tail. The edge whose tail is $v_1$ and head is $v_2$ is distinguished from the edge whose tail is $v_2$ and head is $v_1$. We also allow loops in a directed graph, that is a vertex going from some $v\in V$ to itself.
    The \emph{adjacency matrix} of $A$ of a directed graph $\vec{G} = (V,E)$ is a $|V|\times|V|$ matrix in which $a_{i,j}=1$ if there is an edge in $\vec{G}$ going from $v_i$ to $v_j$. A directed graph is \emph{planar} if it has a drawing in the Euclidean plane without crossing edges.
\end{definition}

There is a bijection between bipartite graphs $G =(U,V,E)$ with $|U| = |V| = n$ and directed graphs $\vec{G} = (V,E)$ on $|V| = n$ vertices. The bijection is that $G = (U,V,E)$ and $\vec{G} = (V,E)$ are images of each other if they have the same adjacency matrix. Note that this bijection does not keep the planarity property. Indeed, the complete bipartite graph $K_{3,3}$ is not planar while the complete directed graph $\vec{G}_3$ has a planar drawing, although both graphs have the same adjacency matrix, the $3\times 3$ all-$1$ matrix.

\subsection{Main results}
\paragraph{History of the problem:}
The main results of this article naturally build on a sequence of important contributions by various authors over the past 40 years. In foundational work, Hartmann~\cite{Hartmann} showed in the 1980s that immanants corresponding to shapes obtained from the partition $(1^n)$ by moving only a constant number of boxes are computable in polynomial time, whereas immanants corresponding to shapes obtained from $(n)$ by moving a constant number of boxes are $\#P$-hard~\cite{Hartmann}. 
Barvinok~\cite{Barvinok} and Bürgisser~\cite{Burgisser-Immanant-complexity} improved the running time for immanants near the determinant; in the same paper Bürgisser also proved $\#P$-hardness for families whose irreducible characters correspond to hook and rectangular partitions. 
Brylinski and Brylinski~\cite{BB2003} established VNP-completeness for partitions whose consecutive parts differ by $\Omega(n^\alpha)$ for some fixed $\alpha>0$. 
Mertens and Moore~\cite{MM2013} proved $\#P$-hardness for immanants associated with partitions containing only $2$’s and $1$’s. 
More recently, de Rugy–Altherre~\cite{dRA} gave $\#P$-hardness for partitions of $n$ with only $n-n^\varepsilon$ parts, each bounded by a constant.

These results suggested the conjecture that every immanant that is $n^{\varepsilon}$-far from the determinant (in the sense that the Ferrers diagram can be obtained from $(n)$ by moving $\Omega(n^{\varepsilon})$ boxes) is $\#P$-hard. The initial aim of this paper was to prove this conjecture for a large fraction of such immanants with $0<\varepsilon$. However, during the preparation of this manuscript, Curticapean~\cite{curticapean} proved a complete dichotomy confirming precisely this phenomenon: computing an immanant is $\#P$-hard whenever the shape is $\Omega(n^\varepsilon)$-far from the determinant for some $\varepsilon>0$, and is otherwise polynomial-time.

After Curticapean’s classification for \emph{general} matrices, the next question is whether hardness persists on \emph{restricted} inputs. 
Indeed, some immanants that are hard in general become easy on structured matrices. 
The most classical example is the permanent: If $A$ is the (possibly weighted) biadjacency matrix of a planar bipartite graph, then $\mathrm{per}(A)$—the number of perfect matchings—can be computed in polynomial time via the FKT/Kasteleyn method (see \cite{ka} or \cite{lp}), whereas computing $\mathrm{per}(A)$ for general $0$-$1$ matrices is already $\#P$-complete~\cite{Valiant-permanent}.%
\footnote{We emphasize our planarity convention. For general immanants, the value is not invariant under independent row/column permutations, so “immanant of a planar bipartite graph’’ is not canonical. 
This is why our restricted-input results are stated for \emph{adjacency matrices of edge-weighted planar directed graphs}, where the matrix is well-defined once the vertex order is fixed. 
The permanent is a special case where row/column permutations do not change the value.}

A second illustration (due to Curticapean~\cite{curticapean}) is the staircase shape $\mu=(k,k-1,\dots,1)$: the $\mu$-immanant of the adjacency matrix of a directed bipartite graph is trivial to compute, because $\chi_\mu(\rho)$ vanishes whenever $\rho$ contains an even cycle, and every cycle in a bipartite graph is even. 
These examples show that the landscape of “matrix class–shape class’’ pairs with accidental tractability can be unexpectedly rich, and motivate the restricted-input hardness results proved here.

\paragraph{Outline of the main contributions:}
In this paper, we study the computational complexity of computing immanants of restricted matrices. In the first part, we restrict the entries of the matrix to $0$s and $1$s and further specialize to adjacency matrices of directed graphs containing only even cycles. Curticapean proved the \#P-hardness of computing the $\lambda$-immanant of $0$-$1$ matrices for those shapes $\lambda$ in which the number of black and white boxes in a checkerboard coloring differ by $\Omega(n^{\varepsilon})$ for some $\varepsilon > 0$. More precisely, he showed that for any shape $\lambda$ with $b(\lambda) = \Omega(n^{\varepsilon})$, $\varepsilon > 0$, either $\Omega(n^{\varepsilon})$ $1 \times 2$ dominos can be peeled off, or after peeling, the remaining staircase shape $\mu = (k, k-1, k-2, \ldots, 1)$ has size $\Omega(n^{\varepsilon})$ (or both). He then proved \#P-hardness for both cases using different graph gadgets. 

For the large remaining staircase case, the gadget proving \#P-hardness is unweighted; that is, the proof holds even when the problem is restricted to $0$-$1$ matrices. In the case of a large domino-tileable part, Curticapean’s proof requires a gadget containing a $-1$ entry. This $-1$ seems essential, as it enables cancellation of terms, and it is unclear how to eliminate it from the construction. Note that the difference in the number of black and white boxes in a staircase shape $\mu$ is $\Omega(k) = \Omega(\sqrt{|\mu|})$. By contrast, any shape that admits a full $1 \times 2$ domino tiling has an equal number of black and white boxes.

In this paper, we prove \#P-hardness for a large class of such domino-tileable shapes. The computation remains hard even when restricted to adjacency matrices of directed graphs in which every cycle has even length. We emphasize that computing the staircase immanant for such matrices is trivial.

More precisely, in Section~\ref{sec:hardness-0-1} we prove the following theorem:

\begin{theorem}\label{theo:hardness-0-1}
Let $\lambda$ be a partition of an even integer $n$ of the form 
\[
\lambda = (w, \mathbf{1} + \lambda_d),
\]
where $|\lambda_d| = n^{\varepsilon}$ for some $0 < \varepsilon$, the shape $\lambda_d$ admits a $1 \times 2$ domino tiling, and 
\[
(3w + 3h(\lambda_d) + 4)\,|\lambda_d| \le n.
\]
Then it is \#P-hard to compute the $\lambda$-immanant of a $0$-$1$ matrix that is the adjacency matrix of a bipartite directed graph. The same hardness result holds for the conjugate partition $\lambda^*$.
\end{theorem}
The proof in Section~\ref{sec:hardness-0-1} is based on a reduction from counting perfect matchings in $3$-regular graphs to computing the indicated immanants. The graph gadget constructed for this proof differs from those used by Curticapean.

Planar bipartite graphs occupy a special place in the study of the computational complexity of immanants. For example, computing the permanent of the adjacency matrix of a planar bipartite graph can be done in polynomial time. This naturally raises the question of which other immanants of adjacency matrices of planar bipartite graphs might also be easy to compute. 

However, there is no canonical way to label the two vertex classes of a bipartite graph. As a consequence, the adjacency matrix of a bipartite graph is not uniquely defined: permuting the vertices within one class corresponds to permuting the rows of the matrix. While row permutations leave the permanent unchanged, they may alter the value of an immanant and even affect the computational complexity of computing it. 

To illustrate this, consider the $1$-regular bipartite graphs on $n+n$ vertices. Depending on the ordering of the vertices in one class, the adjacency matrix $A$ may be any permutation matrix corresponding to some $\rho \in S_n$. In this case,
\[
\mathrm{\Imm}_{\lambda}(A) = \chi_{\lambda}(\rho).
\]
If $\rho$ is the identity permutation, $\chi_{\lambda}(\rho)$ can be computed using the hook-length formula (see \cite{frame,sagan}). In contrast, for a general partition $\lambda$ and permutation $\rho$, even deciding whether $\chi_{\lambda}(\rho)$ is zero is already PP-hard~\cite{ipp}.

The rows of the adjacency matrix of a directed graph cannot be permuted without changing the underlying graph. It is therefore more natural to study the computational complexity of immanants of directed graphs with specified properties. A particularly natural restriction is to consider planar directed graphs. 

In the second part of this paper, we prove that for a large class of shapes, computing the immanant of the weighted adjacency matrix of a planar directed graph is \#P-complete. More precisely, in Section~\ref{sec:hardness-planar-direceted} we prove the following theorem:

\begin{theorem}\label{theo:hardness-weighted-directed-planar}
Let $\lambda = (\mathbf{1}+\lambda_d)$ be a partition of $n$ such that $|\lambda_d| = n^{\varepsilon}$ for some $0 < \varepsilon < \tfrac{1}{2}$. Suppose that for some $w$, the skew shape $\lambda_d/(w)$ can be tiled by $1 \times 2$ dominos. Then it is \#P-hard to compute the $\lambda$-immanant of the adjacency matrix of an edge-weighted planar directed graph. The problem remains \#P-hard even when the edge weights are non-negative integers and $O(n^2)$. %small non-negative integers given in unary.
\end{theorem}

The proof is based on reducing the problem of counting (not necessarily perfect) matchings in planar graphs to computing the indicated immanants. Our construction combines the graph gadget presented in Section~\ref{sec:hardness-0-1} with a gadget from Curticapean’s paper~\cite{curticapean}.

\section{Irreducible characters and the Murnaghan--Nakayama rule}

%(we say it also below) The irreducible characters of $S_n$ are related to the partitions of $n$.
In the following, we give the necessary definitions and theorems to compute irreducible characters. We also state the properties of them  that we use in the \#P-hardness proofs. To be able to compute irreducible characters corresponding to a given shape, we have to define
the notion of border-strip tableaux.

\begin{definition}
Let $\lambda$ and $\rho$ be two partitions of $n$. A \emph{border-strip tableau} of shape $\lambda$ and type $\rho$ is a filling of the Young diagram of shape $\lambda$ with positive integers satisfying:
\begin{enumerate}
    \item The entries are weakly increasing along each row and each column.
    \item For each $i$, the integer $i$ appears exactly $\rho_i$ times.
    \item For each $i$, the squares containing $i$ form a border-strip; that is, a connected skew shape containing no $2 \times 2$ block.
\end{enumerate}
The \emph{height} of a border-strip tableau is the sum of the heights of its border-strips minus the number of strips.
\end{definition}

\begin{example}
Consider the two partitions $\lambda=(5,2,1)$ and $\mu=(3,3,1,1)$. 
Then there are the following six border-strip tableaux of shape $\lambda$:
\[
\begin{array}{@{}c@{\qquad}c@{\qquad}c@{}}
{
 \young(11134,22,2)}
&
{
 \young(11222,13,4)}
&
{
 \young(11222,14,3)}
\\[2.2cm]
{
 \young(12224,13,1)}
&
{
 \young(12223,14,1)}
&
{
 \young(12234,12,1)}
\end{array}
\]
and the corresponding heights are $1,1,1,2,2,3$.
\end{example}

As discussed earlier, the irreducible characters and conjugacy classes of the symmetric group \( S_n \) are naturally indexed by partitions of \( n \). The \emph{Murnaghan--Nakayama rule} provides a purely combinatorial method to compute the value of an irreducible character corresponding to a partition \( \lambda \) on the conjugacy class indexed by a partition \( \rho \). It uses the skew-shapes introduced above and can be stated in a non-recursive form as follows:

\begin{theorem}[Non-recursive Murnaghan--Nakayama rule]
Let \( \lambda \) and \( \rho \) be partitions of \( n \). The irreducible character of \( S_n \) associated to \( \lambda \), evaluated on an element of cycle type \( \rho \), is given by
\begin{equation}
\chi_{\lambda}(\rho) = \sum_{T \in BST(\lambda, \rho)} (-1)^{ht(T)},\label{eq:non-recursive-MN}
\end{equation}
where \( BST(\lambda, \rho) \) denotes the set of border-strip tableaux of shape \( \lambda \) and type \( \rho \), and \( ht(T) \) is the total height of \( T \).
\end{theorem}

The Murnaghan–-Nakayama rule was first proved by Littlewood and Richardson~\cite{LR1934}, building on the Frobenius formula~\cite{Frobenius1900}, and was later reproved independently by Murnaghan~\cite{Murnaghan} and Nakayama~\cite{Nakayama}.

We can say even more about the irreducible characters of $S_n$. For this, we need the notion of $k$-spectrum functions.

\begin{definition}
The \emph{$k$-spectrum} of a permutation $\rho$ is the vector 
\[
(x_1, x_2, \ldots, x_k),
\]
where $x_i$ denotes the number of cycles of length $i$ in $\rho$. The \emph{size} of the $k$-spectrum is 
\[
\sum_{i=1}^k i x_i.
\]

A function $f \colon \mathbb{C}^k \to \mathbb{C}$ is called a \emph{$k$-spectrum function} if it is a polynomial in $k$ variables such that for every monomial $\prod_{i=1}^k x_i^{\beta_i}$ appearing in $f$, 
\[
\sum_{i=1}^{k} i\beta_i \le k.
\]
\end{definition}

Although the following theorem (possibly in slightly different form) is known—for example, it was already used by Hartmann in his paper on the complexity of immanants~\cite{Hartmann}—we include a short proof for completeness.

\begin{theorem}\label{theo:characters-as-k-spectrum-func}
Let $\lambda$ be a partition of $n$ and set $k = n - h(\lambda)$. Then the irreducible character corresponding to $\lambda$ can be written as
\[
\chi_\lambda(\rho) = \mathrm{sign}(\rho)\,
f\bigl(c_1(\rho), c_2(\rho), \ldots, c_k(\rho)\bigr),
\]
where $f$ is a $k$-spectrum function and $(c_1(\rho), c_2(\rho), \ldots, c_k(\rho))$ is the $k$-spectrum of $\rho$. Furthermore, if $\lambda$ and $\lambda'$ are two partitions that differ only in the first column, and $\rho$ and $\rho'$ are two permutations with the same $k$-spectrum, then
\[
\mathrm{sign}(\rho)\,\chi_\lambda(\rho) 
= \mathrm{sign}(\rho')\,\chi_{\lambda'}(\rho').
\]
\end{theorem}

\begin{proof}
Garsia and Goupil~\cite{GarsiaGolupil} proved that the irreducible character corresponding to a partition of the form $(n-|\mu|,\mu)$ is a $|\mu|$-spectrum function in the variables 
\[
c_1(\rho), c_2(\rho), \ldots, c_{|\mu|}(\rho),
\]
and that this function does not depend on the width $w = n - |\lambda_d|$. The first part of the theorem follows from this observation together with the fact that for any partition $\lambda$ and its conjugate $\lambda^*$, one has
\[
\chi_\lambda(\rho) 
= \mathrm{sign}(\rho)\,\chi_{\lambda^*}(\rho)
\]
for all permutations $\rho$~\cite[Equation 2.1.8]{JamesKerber}.
\end{proof}

Theorem~\ref{theo:characters-as-k-spectrum-func} is particularly useful, as it allows the characters corresponding to certain partitions to be evaluated more easily.
\begin{example}
Consider the character $\chi_{(2^{2n})}$ evaluated at the conjugacy class corresponding to $(2^{2n})$, which yields ${2n \choose n}$. 
Computing this value directly using the non-recursive Murnaghan-–Nakayama rule is rather involved. 
However, it becomes straightforward once we observe that
\[
\mathrm{sign}(2^{2n})\,\chi_{(2^{2n})}(2^{2n}) 
= \mathrm{sign}(2n+1,2^{2n})\,\chi_{(2^{2n},1^{2n+1})}(2n+1,2^{2n}).
\]
It is significantly easier to compute $\chi_{(2^{2n},1^{2n+1})}(2n+1,2^{2n})$. 

Any border-strip tableau of shape $(2^{2n},1^{2n+1})$ and type $(2n+1,2^{2n})$ must contain $n$ border-strips of size $2$ in its second column. Indeed, if the border-strip of size $2n+1$ were to include a square from the second column, the remaining skew shape would become disconnected. Moreover, each disconnected component would have an odd number of squares, making it impossible to tile with $2$-border-strips. Thus the strip of size $2n+1$ must occupy the top of the first column, leaving a disconnected remainder consisting of two copies of a $2n\times 1$ shape.

The $n$ strips of size $2$ placed in the second column can be chosen freely from the set $\{1,2,\ldots,2n\}$, and there is a unique way to arrange them: they must be in increasing order from top to bottom. 
Therefore, the total number of such border-strip tableaux is ${2n \choose n}$, and each has the same height.

\end{example}
Surprisingly, it turns out that the parity of the height is a useful invariant for some sets of border-strip tableaux as we show in the following theorem. In fact, this theorem is the key for the hardness result.

\begin{theorem}\label{theo:non-vanishing-characters}
Let $\lambda= (w,\mathbf{1}+\lambda_d)$ be a partition where $\mathbf{1}$ is the all-$1$ vector. Furthermore, assume that  $\lambda_d$ can be tiled with $1\times 2$ dominos. Let $\rho$ be a permutation of size $|\lambda|$ such that it has exactly $|\lambda_d|/2$ cycles of length $2$, it does not have any fixed point, and all other cycles has a length at least $w+h(\lambda')$. Then $\chi_\lambda(\rho)$ is a non-vanishing character.
\end{theorem}
\begin{proof}
The only way to tile $\lambda$ with appropriate border-strips proceeds in the following way.  The border--strip corresponding to the largest cycle of $\rho$ must occupy a hook in the top-left corner of the shape; all other long cycles are placed in the first column, while the border-strips the cycles of length $2$ tile the subshape $\lambda_d$.

By the conditions, at least one such tiling exists. There might be multiple tilings, and thus border-strip tableaux. The sum in the equation~\eqref{eq:non-recursive-MN} could be $0$ if different border-strip tableaux could have different parities of their height, however, the parity of the height is invariant. Indeed, the parity depends on only the number of horizontal $1\times 2$ dominos tiling $\lambda_d$. However, any such domino tiles exactly one square in every second column in $\lambda_d$ while each vertical domino tiles an even number of squares in every second row. Thus, the parity of the number of horizontal dominos is the parity of the sum of the squares in every second row in $\lambda_d$.
\end{proof}

Another important consequence of the Murnaghan--Nakayama rule is that it allows one to combinatorially describe sufficient conditions to decide when a given character has to vanish on certain permutations.
\begin{example}
Consider the partition $\lambda =(2^2,1^{n-4})$.  Then, in fact, $\chi_\lambda(\rho)= 0$ for any permutation in which each cycle has length at least $3$. Indeed, any border-strip tableaux should be filled in with  more than two $1$s. Then in the second column of the shape, $1$ or $2$ squares remain empty while the first two squares in the first column are filled with $1$s. Therefore, a border-strip of size $1$ or $2$ should be filled with a number, but there is no such short cycle in $\rho$. That is, there is no border-strip tableau of shape $\lambda$ and type $\rho$, and thus, the sum in equation~\ref{eq:non-recursive-MN} is empty.
\end{example}
We generalize this observation and put into the theorem that some of the characters are vanishing for the shape $(w,\mathbf{1}+\lambda_d)$.
\begin{theorem}\label{theo:vanishing-characters}
Let $\lambda= (w,\mathbf{1}+\lambda_d)$ be a partition, and let $\rho$ be a permutation such that the size of its $|\lambda_d|+w-1$-spectrum is less than $|\lambda_d|$. Then 
$
\chi_\lambda(\rho)=0.
$
\end{theorem}
\begin{proof}
Without loss of generality, we may assume that the longest cycle of $\rho$ is larger than $w+h(\lambda_d)$, based on Theorem~\ref{theo:characters-as-k-spectrum-func}. Then the border-strip containing $1$s can cover at most $w-1$ of the squares not in the first column. The remaining at least $|\lambda_d|$ squares should be covered with border-strips each of them covering at most $|\lambda_d|+w-1$ squares. However, there are too few such border-strips. That is, there is no border-strip tableau of shape $(w,\mathbf{1}+\lambda_d)$ and type $\rho$, therefore, the sum in equation~\ref{eq:non-recursive-MN} is empty and thus $0$.
\end{proof}

\section{\#P-hardness of computing immanants of $0$-$1$ matrices}
\label{sec:hardness-0-1}

%\begin{definition}
%Let $\lambda$ be a shape of size $n$ and let $A$ be an $n\times n$ matrix. Then the $\lambda$-Immanant of $A$ is defined as
%$$
%\Imm_\lambda(A) := \sum_{\rho\in S_n}\chi_\lambda(\rho) \prod_{i=1}^n %a_{i,\rho(i)}.
%$$
%\end{definition}
In this section we prove the $\#P$-hardness result stated in Theorem~\ref{theo:hardness-0-1}. 
Our approach is to reduce the problem of computing the $\lambda$-immanant of adjacency matrices of directed graphs to the problem of counting perfect matchings in $3$-regular bipartite graphs. 
The latter problem is known to be $\#P$-complete, as shown by Dagum and Luby~\cite{DagumLuby}. 
To carry out the reduction, we introduce a construction that transforms a given $3$-regular bipartite graph into a directed graph with carefully controlled cycle structure.

The key idea is to replace each vertex of the original bipartite graph by a small directed gadget and to connect these gadgets in a way that mirrors the adjacency pattern of the original graph while maintaining strict control over path lengths and cycle parity. 
The resulting directed graph $\vec{G}$ preserves the essential combinatorial information needed for the reduction and ensures that all cycles are even. 

\begin{figure}
    \centering
     \begin{adjustbox}{max totalsize={1\textwidth}{.3\textheight},center}
\begin{tikzpicture}

\draw [black,fill=black] (6.5,5) circle (0.05cm);
\node[] at (6.5,5.5) {$v_c$};

\draw [black,fill=black] (0,5) circle (0.05cm);
\node[] at (-0.5,5) {$v_1$};
\node[] at (1,6) {$v_1^+$};
\node[] at (1,4) {$v_1^-$};

\draw [black,fill=black] (-2,5) circle (0.05cm);
\draw [->,>=triangle 60] (0,5) to [bend right=45] (-2,5);
\draw [->,>=triangle 60] (-2,5) to [bend right=45] (0,5);
%\draw [->,>=triangle 60] (0,10) to [bend right=5] (5,5);
%\draw [->,>=triangle 60] (5,5) to [bend right=5] (0,10);
\draw [black,fill=black] (1,5.5) circle (0.05cm);
\draw [->,>=triangle 60] (0,5) to (1,5.5);
\draw [->,>=triangle 60] (6.5,5) to (1,5.5);
\draw [black,fill=black] (2,6) circle (0.05cm);
\draw [->,>=triangle 60] (1,5.5) to (2,6);
\draw [->,>=triangle 60] (2,6) to (3,6.5);
\node[ rotate=30] at (5,7.5) {$\ldots$};
\draw [->,>=triangle 60] (7,8.5) to (8,9);
\draw [black,fill=black] (8,9) circle (0.05cm);
\draw [->,>=triangle 60] (8,9) to (9,9.5);
\draw [black,fill=black] (9,9.5) circle (0.05cm);
\draw [->,>=triangle 60] (9,9.5) to (10,10);
\draw [->,>=triangle 60] (9,9.5) to (6.5,5);

\draw [black,fill=black] (10,10) circle (0.1cm);
\node[] at (10.5,10.5) {$v_2$};
\node[] at (9,10) {$v_2^-$};
\node[] at (10.5,9) {$v_2^+$};

\draw [black,fill=black] (12,12) circle (0.05cm);
\draw [->,>=triangle 60] (10,10) to [bend right=45] (12,12);
\draw [->,>=triangle 60] (12,12) to [bend right=45] (10,10);
%\draw [->,>=triangle 60] (10,10) to [bend right=5] (5,5);
%\draw [->,>=triangle 60] (5,5) to [bend right=5] (10,10);
\draw [black,fill=black] (10,9) circle (0.05cm);
\draw [->,>=triangle 60] (10,10) to (10,9);
\draw [->,>=triangle 60] (6.5,5) to (10,9);
\draw [black,fill=black] (10,8) circle (0.05cm);
\draw [->,>=triangle 60] (10,9) to (10,8);
\draw [->,>=triangle 60] (10,8) to (10,7);
\node[] at (10,6) {$\vdots$};
\draw [->,>=triangle 60] (10,3) to (10,2);
\draw [black,fill=black] (10,2) circle (0.05cm);
\draw [->,>=triangle 60] (10,2) to (10,1);
\draw [black,fill=black] (10,1) circle (0.05cm);
\draw [->,>=triangle 60] (10,1) to (10,0);
\draw [->,>=triangle 60] (10,1) to (6.5,5);

\draw [black,fill=black] (10,0) circle (0.1cm);
\node[] at (10.5,-0.5) {$v_3$};
\node[] at (9,0) {$v_3^+$};
\node[] at (10.5,1) {$v_3^-$};

\draw [black,fill=black] (12,-2) circle (0.05cm);
\draw [->,>=triangle 60] (10,0) to [bend right=45] (12,-2);
\draw [->,>=triangle 60] (12,-2) to [bend right=45] (10,0);
%\draw [->,>=triangle 60] (10,0) to [bend right=5] (5,5);
%\draw [->,>=triangle 60] (5,5) to [bend right=5] (10,0);
\draw [black,fill=black] (9,0.5) circle (0.05cm);
\draw [->,>=triangle 60] (10,0) to (9,0.5);
\draw [->,>=triangle 60] (6.5,5) to (9,0.5);
\draw [black,fill=black] (8,1) circle (0.05cm);
\draw [->,>=triangle 60] (9,0.5) to (8,1);
\draw [->,>=triangle 60] (8,1) to (7,1.5);
\node[rotate=-30] at (5,2.5) {$\ldots$};
\draw [->,>=triangle 60] (3,3.5) to (2,4);
\draw [black,fill=black] (2,4) circle (0.05cm);
\draw [->,>=triangle 60] (2,4) to (1,4.5);
\draw [black,fill=black] (1,4.5) circle (0.05cm);
\draw [->,>=triangle 60] (1,4.5) to (0,5);
\draw [->,>=triangle 60] (1,4.5) to (6.5,5);

\end{tikzpicture}%
\end{adjustbox}

    \caption{The graph gadget replacing each vertex in a $3$-regular graph. See Construction \ref{const:1} for details.}
    \label{fig:gadget}
\end{figure}
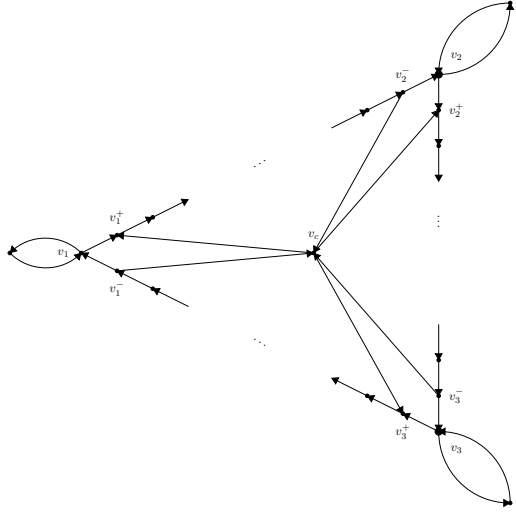

\begin{const}\label{const:1}
Let \( H = (V, E) \) be a \( 3 \)-regular bipartite graph with \( |\lambda_d|/2 \) vertices in each part of the bipartition. 
Replace each vertex \( v \in V \) with the gadget shown in Figure~\ref{fig:gadget}. 
Each gadget contains the distinguished vertices
\[
v_c, \;
v_1^-, v_1, v_1^+, \;
v_2^-, v_2, v_2^+, \;
v_3^-, v_3, v_3^+.
\]

\smallskip
\emph{Internal edges.}
For each \( i = 1, 2, 3 \), add the directed edges
\[
(v_i^-, v_i),\;
(v_i, v_i^+),\;
(v_i^-, v_c),\;
(v_c, v_i^+).
\]

\smallskip
\emph{Connecting paths.}
For \( i = 1, 2 \), insert a path of odd length
\[
p = w + h(\lambda_d) - 1
\quad \text{or} \quad
p = w + h(\lambda_d),
\]
choosing the odd value, from \( v_i^+ \) to \( v_{i+1}^- \).
Also include a path of length \( p \) from \( v_3^+ \) to \( v_1^- \).

\smallskip
\emph{Inter-gadget connections.}
For each edge \( (v, w) \in E \), suppose \( w \) is the \( i^\text{th} \) neighbor of \( v \) and \( v \) is the \( j^\text{th} \) neighbor of \( w \). 
Introduce the directed edges
\[
(v_i, w_j) 
\quad \text{and} \quad 
(w_j, v_i).
\]

\smallskip
\emph{Adjusting the size.}
The number of vertices created so far satisfies
\[
\le (3p + 4)\, |\lambda_d|
= (3w + 3h(\lambda_d) + 4)\, |\lambda_d|.
\]
If this is smaller than \( n \), elongate exactly one of the paths of length \( p \) in a single gadget so that the total number of vertices becomes \( n \).  No further vertices or edges are added and the resulting directed graph is denoted by \( \vec{G} \).
\
\end{const}

\begin{remark}
    Since \( \lambda_d \) is domino-tileable, \( |\lambda_d| \) is even. Therefore also 
    \[
    n - (3w + 3h(\lambda_d) + 4)\, |\lambda_d|
    \]
    is also even, ensuring that the elongated path preserves its odd length. Furthermore, notice that every cycle in \( \vec{G} \) has even length: Indeed, cycles entirely within a gadget are even by construction. Furthermore, since \( H \) is bipartite, any cycle that spans multiple gadgets is also even.
\end{remark}

This gadget construction will be the central tool for our reduction, ensuring both the correct vertex count and the crucial even-cycle property required in the proof.

The following properties of the directed graph \( \vec{G} \) constructed in Section~\ref{const:1} will be crucial.

\begin{lemma}\label{lemmaconst}
The directed graph \( \vec{G} \) obtained from Construction~\ref{const:1} satisfies:
\begin{enumerate}
    \item Every cycle of length \( 2 \) in any cycle cover of \( \vec{G} \) consists of a pair of vertices \( (v_i, w_j) \) corresponding to an edge \( (v, w) \in E(H) \).
    
    \item Any cycle in a cycle cover of \( \vec{G} \) that is longer than \( 2 \) has length at least 
    \[
    p + 1 \ge w + h(\lambda_d).
    \]
    
    \item There is a one-to-one correspondence between perfect matchings of \( H \) and cycle covers of \( \vec{G} \) containing exactly \( |V|/2 \) cycles of length \( 2 \).
    
    \item No cycle cover of \( \vec{G} \) contains more than \( |V|/2 \) cycles of length \( 2 \).
\end{enumerate}
\end{lemma}

\begin{proof}
\begin{enumerate}
    \item By construction, the only \( 2 \)-cycles in \( \vec{G} \) are those formed by pairs \( (v_i, w_j) \) for \( (v, w) \in E(H) \). No other \( 2 \)-cycles exist.
    
    \item A cycle that is not a \( 2 \)-cycle must leave a gadget through some \( v_i^+ \) and return via a connecting path. By construction, any such path has length at least \( p \), so the entire cycle has length at least \( p + 1 \ge w + h(\lambda_d) \).
    
    \item Let \( M \subseteq E(H) \) be a perfect matching. For each \( (v, w) \in M \), include the \( 2 \)-cycle on vertices \( (v_i, w_j) \) determined by the local indices. Within each gadget corresponding to \( v \), the remaining vertices are uniquely covered by a single directed cycle once \( v_i \) is occupied by a \( 2 \)-cycle. This produces a unique cycle cover of \( \vec{G} \) containing exactly \( |V|/2 \) cycles of length \( 2 \).
    
    Conversely, suppose a cycle cover of \( \vec{G} \) contains \( |V|/2 \) cycles of length \( 2 \). Since no gadget can contribute more than one \( 2 \)-cycle (otherwise some \( v_i^+ \) could not be covered), each such cycle corresponds to a distinct edge of \( H \), and these edges form a perfect matching. Thus there is a bijection between perfect matchings of \( H \) and cycle covers of \( \vec{G} \) with \( |V|/2 \) \( 2 \)-cycles.
    
    \item The previous argument shows that each gadget can contribute at most one \( 2 \)-cycle, so the total number of such cycles in any cover is bounded by the number of vertices in one side of the bipartition, namely \( |V|/2 \).
\end{enumerate}
\end{proof}

With the properties of cycle covers of $\vec{G}$ established above we can show the \#P-hardness.
\begin{proof}[Proof of Theorem~\ref{theo:hardness-0-1}]

Let $A$ be the adjacency matrix of the graph $\vec{G}$ resulting from Construction $\ref{const:1}$ from a 3-regular bipartite graph $H$, which has $|\lambda_d|/2$ many vertices in both vertex classes. Consider the corresponding immanant
$$
\Imm_\lambda(A) = \sum_{\rho\in S_n} \chi_\lambda(\rho) \prod_{i=1}^n a_{i,\rho(i)}.
$$
First, we observe which of the summands  are actually vanishing: 
If $\rho$ is not a cycle cover in $\vec{G}$, then $\prod_{i=1}^n a_{i,\rho(i)} = 0$. Furthermore, if $\rho$ is a cycle cover, but does not contain $|\lambda_d|/2$ $2$-cycles, then it contains less than $|V|/2$ cycles of length $2$, and thus $\Imm_\lambda(\rho) = 0$, according to Theorem~\ref{theo:vanishing-characters}. Indeed, observe that any cycle $\rho$ which is not a $2$-cycle has size $p+1$, and $p+1\ge w+|\lambda_d|$.

Finally, if $\rho$ is a cycle cover in $\vec{G}$ and contains $|\lambda_d|/2$ $2$-cycles, then by Lemma \ref{lemmaconst} (3) it corresponds to a perfect matching in $H$, therefore, we get that
$$
\Imm_\lambda(A) = \chi_\lambda(n-|\lambda_d|,2^{|\lambda_d|/2}) PM(H), 
$$
where $PM(H)$ is the number of perfect matchings in $H$. 
Since  $\chi_\lambda(n-|\lambda_d|,2^{|\lambda_d|/2})$ is a non-vanishing character, according to Theorem~\ref{theo:non-vanishing-characters} we obtain
$$
PM(H) = \frac{\Imm_\lambda(A)}{\chi_\lambda(n-|\lambda_d|,2^{|\lambda_d|/2})}.
$$
Note that by construction the size of $A$ is polynomial in the number of vertices of $H$
Exhibit a directed graph $\vec{G}_1$ consists of a directed cycle of length $n-|\lambda_d|$ and $2^{|\lambda_d|/2}$ disjoint cycles of length $2$. Then for its adjacency matrix $A_1$, we have that $\Imm_\lambda(A_1) = \chi_\lambda(n-|\lambda_d|,2^{|\lambda_d|/2})$. That is,
$$
PM(H) = \frac{\Imm_\lambda(A)}{\Imm_\lambda(A_1)}.
$$
From the orthonormality relation of the irreducible characters, any immanant of a $0$-$1$ matrix is at most $n!$ (see, for example, \cite{sagan}, Theorem 1.9.3). So it follows that the logarithm of any immanant of a $0$-$1$ matrix is upper bounded by a polynomial of the size of the matrix. Therefore, $\Imm_\lambda(A)$ and $\Imm_\lambda(A_1)$ contains at most polynomially many digits. Therefore, the ratio of $\Imm_\lambda(A)$ and $\Imm_\lambda(A_1)$ can also be computed in polynomial time.
We can conclude that any algorithm computing $\Imm_\lambda(A)$  and $\Imm_\lambda(A_1)$ in polynomial time is applicable to compute $PM(H)$ in polynomial time. Recall that $H$ is a general $3$-regular bipartite graph, and computing $PM(H)$ is \#P-complete \cite{DagumLuby}. Thus, computing $\Imm_{\lambda}$ of a $0$-$1$ matrix is also \#P-complete.  
%Thus any algorithm computing $\Imm_\lambda(A)$  in polynomial time is applicable to compute $PM(H)$ in polynomial time since $A$ has size which is polynomial in the size of $H$, and, furthermore, since $\chi_\lambda(n-|\lambda_d|,2^{|\lambda_d|/2})$ can be computed in polynomial time. 

Finally, in order to show that the theorem also holds for $\lambda^*$, observe that the immanant is non-vanishing only on one particular cycle structure.% the theorem holds also for $\lambda^*$. 
\end{proof}

There is an upper limit how large $\lambda_d$ might be. Clearly, if $w = O(|\lambda_d|)$, then $\varepsilon$  might be any number smaller than $\frac{1}{2}$. If $w = O(\sqrt{|\lambda_d|})$ and $\lambda_d$ has a square shape, then $\varepsilon$ might be any number smaller than $\frac{2}{3}$.

\section{\#P-hardness result on computing immanants of adjacency matrices of edge-weighted, planar, directed graphs}
\label{sec:hardness-planar-direceted}
Finally, we prove the hardness result stated in Theorem \ref{theo:hardness-weighted-directed-planar}  by reducing the number of (not necessarily perfect) matchings of planar graphs to computing the $\lambda$-immanant of the adjacency matrix of edge-weighted, planar, directed graphs. Computing the number of (not necessarily perfect) matchings in general, unweighted planar graphs is \#P-complete \cite{jerrum-monomer-dimer,jerrum-erratum}. 
\begin{const}\label{const:2}
Let $H = (V,E)$ be a general, unweighted planar graph on $\frac{n^{\varepsilon}}{2}$ vertices. We construct a corresponding directed, planar graph $\vec{G}$ in the following way. Replace each edge $(u,v)\in E$ with the gadget shown in Figure~\ref{fig:radus-gadget}. We call this gadget the \emph{match gadget}.
The indicated edge has weight $-1$, all other edges have weight $1$. For each vertex $v\in H$, add a gadget represented in Figure~\ref{fig:triangle-gadget}. The indicated edge has a weight $x$, all other edges have weight $1$. The paths between $u_1^+$ to $u_2^-$ as well as from $u_2^+$ to $u_1^-$, from $w_1^+$ to $w_2^-$ and from $w_2^+$ to $w_1^-$ are all have length $p$ such that $(4p+7)|V| + 2 |V|+ 2|E|=n$. That is,
$$
p = \frac{n- 2|E|-9|V|}{4|V|}.
$$
Note that in any planar graph, $|E| \le 3|V|-6$, therefore
$$
p \ge \frac{n^{1-\varepsilon}}{2} -\frac{15}{4} - \frac{12}{n^{\varepsilon}}.
$$
We call this gadget the \emph{vertex covering gadget}.

Finally, we add $2|V|$ additional vertices to $\vec{G}$ and create $|V|$ isolated $2$-cycles on them. There are no more vertices and edges in $\vec{G}$.

\end{const}
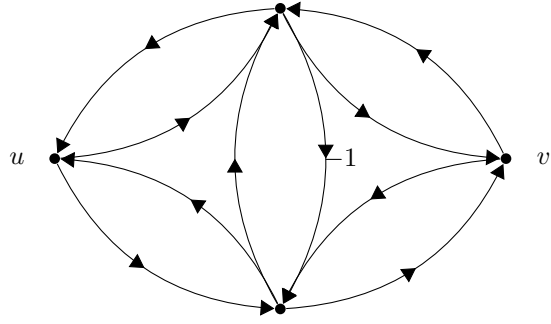
\begin{figure}
    \centering
     \begin{adjustbox}{max totalsize={1\textwidth}{.5\textheight},center}
\begin{tikzpicture}
%\tikzset{vertex/.style = {shape=circle,draw,minimum size=0.05cm}}
\tikzset{vertex/.style = {shape = circle,fill = black,minimum size = 0.05cm,inner sep=0.05cm}}
%\tikzset{edge/.style = {->,>=triangle 60}}
\tikzset{edge/.style={->,>=triangle 60,decoration={
  markings,
  mark=at position .5 with {\arrow{>}}},postaction={decorate}}}

%\draw [black,fill=black] (0,5) circle (0.05cm);
\node[] at (-0.5,5) {$u$};
\node[vertex] (c) at  (0,5) {};
%\draw [->,>=triangle 60] (0,5) to [bend right=20] (3,7);
%\draw [->,>=triangle 60] (3,7) to [bend right=20] (0,5);
%\draw [black,fill=black] (3,7) circle (0.05cm);

\node[vertex] (a) at  (3,7) {};

%\draw [->,>=triangle 60] (0,5) to [bend right=20] (3,3);
%\draw [->,>=triangle 60] (3,3) to [bend right=20] (0,5);
%\draw [black,fill=black] (3,3) circle (0.05cm);
\node[vertex] (b) at  (3,3) {};

%\draw [->,>=triangle 60] (3,7) to [bend right=20] (3,3);
\draw[edge] (b)  to[bend left] (a);
\node[] at (3.8,5) {$-1$};
%\draw [->,>=triangle 60] (3,3) to [bend right=20] (3,7);
\draw[edge] (a)  to[bend left] (b);

%\draw [->,>=triangle 60] (3,7) to [bend right=20] (6,5);
%\draw [->,>=triangle 60] (6,5) to [bend right=20] (3,7);

%\draw [->,>=triangle 60] (3,3) to [bend right=20] (6,5);
%\draw [->,>=triangle 60] (6,5) to [bend right=20] (3,3);

%\draw [black,fill=black] (6,5) circle (0.05cm);
\node[vertex] (d) at  (6,5) {};
\node[] at (6.5,5) {$v$};
\draw[edge] (b)  to[bend right] (c);
\draw[edge] (c)  to[bend right] (b);
\draw[edge] (a)  to[bend right] (d);
\draw[edge] (d)  to[bend right] (a);
\draw[edge] (b)  to[bend right] (d);
\draw[edge] (d)  to[bend right] (b);
\draw[edge] (c)  to[bend right] (a);
\draw[edge] (a)  to[bend right] (c);

\end{tikzpicture}%
\end{adjustbox}

    \caption{The match gadget replacing each edge in a planar graph. See the Construction \ref{const:2} for details} 
    \label{fig:radus-gadget}
\end{figure}
Note that the gadget in Figure~\ref{fig:radus-gadget} is identical with the gadget in \cite{curticapean}. Here we state properties of this match gadget that are not hard to verify. Detailed proofs of these statements can be found in the cited manuscript.
\begin{figure}
    \centering
     \begin{adjustbox}{max totalsize={1\textwidth}{.5\textheight},center}
\begin{tikzpicture}

\draw [black,fill=black] (7,1) circle (0.05cm);
\node[] at (7,0.5) {$v$};

\draw [->,>=triangle 60] (7,1) to [bend right=20] (14,5);
\node[] at (12,2.5) {$x$};
\draw [->,>=triangle 60] (0,5) to [bend right=20] (7,1);
\draw [black,fill=black] (14,5) circle (0.05cm);
\node[] at (14.5,5) {$u_1$};
\node[] at (13,6.5) {$u_1^+$};
\node[] at (12.85,4.2) {$u_1^-$};
\draw [black,fill=black] (0,5) circle (0.05cm);
\node[] at (-0.5,5) {$w_2$};
\node[] at (1,6.5) {$w_2^-$};
\node[] at (1.15,4.2) {$w_2^+$};

\draw [->,>=triangle 60] (14,5) to (13,6);
\draw [black,fill=black] (13,6) circle (0.05cm);
\draw [->,>=triangle 60] (13,6) to (12.5,6);
\draw [black,fill=black] (12.5,6) circle (0.05cm);
\draw [->,>=triangle 60] (12.5,6) to (12,6);

\node[] at (10.5,6) {$\ldots$};
\draw [->,>=triangle 60] (9.5,6) to (9,6);
\draw [black,fill=black] (9,6) circle (0.05cm);
\draw [->,>=triangle 60] (9,6) to (8.5,6);
\draw [black,fill=black] (8.5,6) circle (0.05cm);
\draw [->,>=triangle 60] (8.5,6) to (8,5.5);
\draw [black,fill=black] (8,5.5) circle (0.05cm);
\draw [->,>=triangle 60] (8,5.5) to (8.5,4.5);
\draw [black,fill=black] (8.5,4.5) circle (0.05cm);
\draw [->,>=triangle 60] (8.5,4.5) to (9,4.5);
\draw [black,fill=black] (9,4.5) circle (0.05cm);
\draw [->,>=triangle 60] (9,4.5) to (9.5,4.5);

\node[] at (10.5,4.5) {$\ldots$};
\draw [->,>=triangle 60] (12,4.5) to (12.5,4.5);
\draw [black,fill=black] (12.5,4.5) circle (0.05cm);
\draw [->,>=triangle 60] (12.5,4.5) to (13,4.5);
\draw [black,fill=black] (13,4.5) circle (0.05cm);
\draw [->,>=triangle 60] (13,4.5) to (14,5);

\draw [->,>=triangle 60] (8,5.5) to [bend right=20] (6,5.5);
\draw [->,>=triangle 60] (6,5.5) to [bend right=20] (8,5.5);
\draw [black,fill=black] (6,5.5) circle (0.05cm);
\node[] at (8.5,5.5) {$u_2$};
\node[] at (8.5,6.5) {$u_2^-$};
\node[] at (8.5,4) {$u_2^+$};
\node[] at (5.5,5.5) {$w_1$};
\node[] at (5.5,6.5) {$w_1^+$};
\node[] at (5.5,4) {$w_1^-$};

\node[] at (2.7,5.35) {$w_c$};
\draw [black,fill=black] (3,5.25) circle (0.05cm);
\draw [->,>=triangle 60] (3,5.25) to (5.5,6);
\draw [->,>=triangle 60] (3,5.25) to (1,4.5);
\draw [->,>=triangle 60] (5.5,4.5) to (3,5.25);

\node[] at (11.3,5.35) {$u_c$};
\draw [black,fill=black] (11,5.25) circle (0.05cm);
\draw [->,>=triangle 60] (8.5,6) to (11,5.25);
\draw [->,>=triangle 60] (11,5.25) to (8.5,4.5);
\draw [->,>=triangle 60] (13,4.5) to (11,5.25);

%%%%%%%%%%%%%%%%%%%

\draw [->,>=triangle 60] (6,5.5) to (5.5,6);
\draw [black,fill=black] (5.5,6) circle (0.05cm);
\draw [->,>=triangle 60] (5.5,6) to (5,6);
\draw [black,fill=black] (5,6) circle (0.05cm);
\draw [->,>=triangle 60] (5,6) to (4.5,6);

\node[] at (3.5,6) {$\ldots$};
\draw [->,>=triangle 60] (2,6) to (1.5,6);
\draw [black,fill=black] (1.5,6) circle (0.05cm);
\draw [->,>=triangle 60] (1.5,6) to (1,6);
\draw [black,fill=black] (1,6) circle (0.05cm);
\draw [->,>=triangle 60] (1,6) to (0,5);

\draw [->,>=triangle 60] (0,5) to (1,4.5);
\draw [black,fill=black] (1,4.5) circle (0.05cm);
\draw [->,>=triangle 60] (1,4.5) to (1.5,4.5);
\draw [black,fill=black] (1.5,4.5) circle (0.05cm);
\draw [->,>=triangle 60] (1.5,4.5) to (2,4.5);

\node[] at (3.5,4.5) {$\ldots$};
\draw [->,>=triangle 60] (4.5,4.5) to (5,4.5);
\draw [black,fill=black] (5,4.5) circle (0.05cm);
\draw [->,>=triangle 60] (5,4.5) to (5.5,4.5);
\draw [black,fill=black] (5.5,4.5) circle (0.05cm);
\draw [->,>=triangle 60] (5.5,4.5) to (6,5.5);

\end{tikzpicture}%
\end{adjustbox}

    \caption{The vertex covering gadget attached to each vertex of a planar graph. See Construction \ref{const:2} for details.}
    \label{fig:triangle-gadget}
\end{figure}
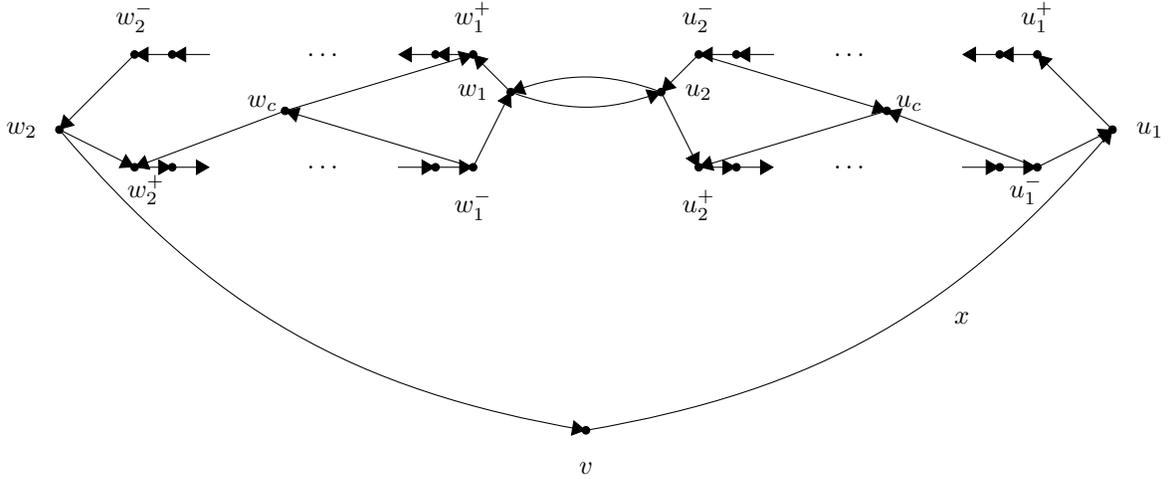
\begin{proposition}
The match gadget shown in Figure~\ref{fig:radus-gadget} has the following properties.
\begin{enumerate}
    \item If both $u$ and $v$ are in a cycle of a cycle cover not involving edges in the match gadget between $u$ and $v$, then the remaining two vertices can be covered in exactly one way, with weight $-1$.
    \item If exactly one of the vertices $u$ and $v$ are in a cycle of a cycle cover not involving edges in the match gadget between $u$ and $v$, then the remaining three vertices in this gadget can be covered in two different ways, and the sum of the weights of these cycle covers cancel each other.
    \item If both $u$ and $v$ are in one or two cycles of a cycle cover involving only edges in the match gadget between $u$ and $v$, then there are four such ways, two of them contain two $2$-cycles, two of them contain one $4$-cycle. In all cases, the weight of the cycle(s) is $1$.
    \item If both $u$ and $v$ are in one cycle of a cycle cover containing both edges in the match gadget between $u$ and $v$ and edges outside this gadget, then such cycles can be paired such that their weights cancel each other and the pair of cycle covers have exactly the same cycle length structures. What follows is that in any immanant evaluated on the adjacency matrix of $\vec{G}$, the cycle covers with any cycle that contains both edges from a match gadget between two vertices and edges outside of that gadget cancel each other.
\end{enumerate}
\end{proposition}
We now examine the connection of matchings in the original graph $H$ to cycle covers in the resulting graph $\vec{G}$:

Consider a vertex $v\in H$ and its corresponding vertex $v\in \vec{G}$. If $v$ is covered with a cycle containing edges from the vertex covering gadget in Figure~\ref{fig:triangle-gadget}, then there is one way for it, furthermore, the remaining vertices in the vertex covering gadget can be covered in exactly one way: with a cycle containing vertices $w_c$, $w_2^+$, $w_1^-$ and a cycle containing vertices $u_c$, $u_2^+$, $u_1^-$. The product of the weights of these cycles is $x$.

If $v$ is covered with a cycle not containing edges from its vertex covering gadget, then the remaining vertices in the vertex covering gadget of $v$ can be covered in exactly one way, with three cycles: the first is the cycle between vertices $u_2$ and $w_1$, the second is the cycle that contains the vertices $w_c$, $w_1^+$, $w_1^-$ and the third is the cycle that contains the vertices $u_c$, $u_2^+$, $u_2^-$.

Based on these observations we have the following.

\begin{proposition}
For the pair of graphs $H$ and $\vec{G}$ in Construction \ref{const:2}  
there is a one-to-many correspondence between the matchings of $H$ and the cycle covers of $\vec{G}$.  Moreover, to a matching in $H$ containing  $k$ edges,  there are $4^k$ corresponding cycle covers in $\vec{G}$. Each of these has a weight $x^{|V|-2k}(-1)^{|E|-k}$, and for each $m = 0, 1, \ldots, k$ exactly $2^k{k\choose m}$  of them have the following cycle structure: 
\begin{enumerate}
    \item $2(k-m) + |E|-k + 2k + |V|$ cycles of length $2$: two cycles in each of the $k-m$ match gadgets corresponding to edges in the matching of $H$,  one cycle in each of the $|E| -k$ match gadgets corresponding to edges not in the match of $H$, one cycle in  each of the $2k$ vertex cover gadgets corresponding to vertices incident to edges in the matching of $H$ and the additional $|V|$ $2$-cycles,
    \item $m$ cycles of length $4$ (one cycle in each of the $m$ match gadgets corresponding to edges in the matching of $H$),
    \item $4k$ cycles of length $2p+2$ (two cycles in  each of the $2k$ vertex cover gadgets corresponding to vertices incident to edges in the matching of $H$),
    \item $|V|-2k$ cycles of length $2p+5$ (one cycle in each of the $|V|-2k$ vertex cover gadgets corresponding to vertices not incident to edges in the matching of $H$)
    \item and $2|V|-4k$ cycles of length $p+1$ (two cycles in each of the $|V|-2k$ vertex cover gadgets corresponding to vertices not incident to edges in the matching of $H$).
\end{enumerate}
\end{proposition}

 As a cross check, the number of vertices in such a cycle cover is
\begin{eqnarray*}
4m + 2(2(k-m) + |E| -k + 2k +|V|) + (2p+5)(|V|-2k) + (p+1)(2|V|-4k) + (2p+2)4k\\ = (4p+7)|V| + 2|V| + 2|E|,
\end{eqnarray*}
which is indeed the number of vertices in $\vec{G}$.
%\end{proof}
Note that by construction one of the weights in the  constructed directed graph $\vec{G}$ is $x$ and thus, we can view its adjacency matrix as a linear matrix polynomial in $x$. This in particular yields that also the immanant is a polynomial in $x$. More concretely: 
\begin{proposition}\label{lempoly}
Let $H=(V,E)$ be a planar graph on $\frac{n^\epsilon}{2}$ many vertices, let $A(\vec{G})$ denote the adjacency matrix of the directed graph $\vec{G}$ obtained from $H$ by Construction \ref{const:2}.
Then, the immanant $\Imm_\lambda(A(\vec{G}))$ is a polynomial in $x$ of degree ${\lfloor|V|/2\rfloor}$. More concretely, we have
\begin{equation}\label{eq:ima}
    \Imm_\lambda(A(\vec{G})) = \sum_{k=0}^{\lfloor|V|/2\rfloor} M(H,k) x^{|V|-2k} (-1)^{|E|-k} 2^k  \sum_{m=0}^k {k\choose m}\chi_\lambda(\rho(|E|,|V|,k,m)),
\end{equation}
where $M(H,k)$ denotes  the number of $k$-matchings in $H$ and  
\[\rho(|E|,|V|,k,m) \text{ is }    ((2p+5)^{|V|-2k},(2p+2)^{4k},(p+1)^{2|V|-4k},4^{m},2^{2(k-m) + |E|-k + 2k +|V|})
\]
\end{proposition}
With these preparations we are able to give the proof of the \#P hardness for edge-weighted,  planar,  directed  graphs.
\begin{proof}[Proof of Theorem~\ref{theo:hardness-weighted-directed-planar}]
%We prove the hardness by reducing the number of (not necessarily perfect) matchings of planar graphs to computing the $\lambda$-Immanant of the adjacency matrix of edge-weighted, planar, directed graphs. Computing the number of matchings in planar graphs is \#P-complete \cite{jerrum-monomer-dimer,jerrum-erratum}. First we give the proof for edge-weighted graphs with possible $-1$ weights then we show how to eliminate the $-1$ weights.
First we give the proof for edge-weighted graphs with possible $-1$ weights then we show how to eliminate the $-1$ weights.

We will use the gadget described in  Construction~\ref{const:2} to show  that a polynomial time algorithm to compute the immanant would also yield a polynomial time algorithm to calculate for all $k$ the number of $k$-matchings in a planar graph $H$. To see this, we  observe first that according to  Proposition \ref{lempoly} the immanant of the adjacency matrix of $A(\vec{G})$  %$\Imm_\lambda(A(\vec{G}))$ 
is  a polynomial in $x$ of degree  $\lfloor|V|/2\rfloor$. 
In order to obtain the corresponding coefficients we can use interpolation: Assigning $\lfloor|V|/2\rfloor+1$ different values to $x$ 
and computing the immanant of the resulting matrix we have access to all of the coefficients. Therefore, if $\Imm_\lambda(A(\vec{G}))$ is computable in a polynomial time we can also compute all of these coefficients in polynomial time.

In order to use Equation \eqref{eq:ima} to obtain the $k$-matchings in $H$ we have to make sure that the sum
\begin{equation}\label{eq:non-vanishing-sum-of-chars}
 \sum_{m=0}^k {k\choose m}\chi_\lambda(\rho(|E|,|V|,k,m)   
\end{equation}
does not vanish and can be computed in polynomial time. Indeed, we know from Theorem~\ref{theo:characters-as-k-spectrum-func} that $\chi_\lambda$ is a $b(\lambda)$-spectrum function. Since $b(\lambda) < p+1$ we have that
$$
\chi_\lambda(\rho(|E|,|V|,k,m)) = (-1)^{|V|}\chi_{\lambda'}(4^{m},2^{2(k-m) + |E|-k + 2k +|V|})
$$
where $\lambda'$ obtained from $\lambda$ by deleting $((2p+5)(|V|-2k)+4k(2p+2)+(p+1)(2|V|-4k)$ squares from its first row.
For such characters 
Curticapean proved that
$$
\sum_{k=0}^m {k\choose m} \chi_{\lambda'}(4^m,2^{2(k-m) + |E|-k + 2k +|V|})
$$
is not vanishing and can be computed in polynomial time \cite{curticapean}, thus the value in equation~\eqref{eq:non-vanishing-sum-of-chars} is not vanishing and also computable in polynomial time. Therefore, for each $k$, $M(H,k)$ could be computable in polynomial time if $\Imm_\lambda(A(\vec{G}))$ could be computed in polynomial time. Indeed,
$$
M(H,k) = \frac{\alpha_{|V|-2k}}{(-1)^{|E|-k}2^k\sum_{m=0}^k {k\choose m} \chi_\lambda(\rho(|E|,|V|,k,m))}
$$
where $\alpha_{|V|-2k}$ is the coefficient of $x^{|V|-2k}$ in $poly(x)$ as $\Imm_\lambda(A(\vec{G}))$.

The $-1$ weights can be eliminated if for some appropriate sets of prime numbers $p = p_1, p_2, \ldots, p_s$, the $-1$ weight is replaced with $p-1$, and then each $M(H,k)$ is computed modulo $p$. Using the Chinese Remainder Theorem, $M(H,k)$ can be computed modulo $\prod_{i=1}^s p_i$. Since $M(H,k)$ is definitely smaller than $2^{|E|} \le 2^{3|V|-6}$, it provides an exact computation if $\prod_{i=1}^s p_i > 2^{3|V|-6}$. The appropriate prime numbers are those that are odd and does not divide any of the coefficients
$$
\sum_{m=0}^k {k\choose m} \chi_\lambda(\rho(|E|,|V|,k,m)).
$$
It is definitely possible to select such prime numbers from the first prime numbers whose product is larger than
$$
2\times 2^{3|V|-6}\prod_{k=0}^{|V|/2}\left(\sum_{m=0}^k {k\choose m} \chi_\lambda(\rho(|E|,|V|,k,m))\right).
$$
A very crude approximation
$$
\sum_{m=0}^k {k\choose m} \chi_\lambda(\rho(|E|,|V|,k,m) \le 2^k n^{n^{\varepsilon}} \le 2^k n^{\sqrt{n}}
$$
suffices. To see this inequality, observe that in any border-strip tableaux, the $\lambda_d$ part must be filled in with numbers varying between $1$ and $n$ and ${k\choose m} \le 2^k$. Then we are looking for a $t$ such that the product of the prime numbers at most $t$ be at least
$$
2\times 2^{3|V|-6} \prod_{k=0}^{|V|/2} 2^k n^{\sqrt{n}} \le 2^{3n^{\varepsilon} + \frac{n^{\varepsilon}(n^{\varepsilon}-1)}{2}} n^{n^{\varepsilon}\sqrt{n}} \le 2^{3\sqrt{n}+ n+\log_2(n)n} \le 2^{n^2}.
$$
We know that the product of the prime numbers smaller than $t$ is at least $2^{t/2}$ \cite{Erdos}. Then $t=2n^2$ suffices. It is possible to find all prime numbers below $2n^2$ in polynomial time, and it is also possible in polynomial time to select those that do not divide $\sum_{m=0}^k {k\choose m} \chi_\lambda(\rho(|E|,|V|,k,m)$ for any $k$, compute the corresponding immanants and solve exactly $M(H,k)$ for each $k$ using the Chinese Remainder Theorem. 
\end{proof}

We would like to mention that extending Theorem~\ref{theo:hardness-weighted-directed-planar} to $0$-$1$ matrices has potential technical difficulties. A possible idea is to use Valiant's approach to replace a directed edge $e = (u,v)$ in a graph $G = (V,E)$ with small natural number weight $x$ by a gadget of size $O(x)$ without edge weights \cite{Valiant-permanent}. In the so-obtained graph $G'$, there is exactly one cycle cover covering all of the vertices except $u$ and $v$ (corresponding to cycle covers in $G$ in which $u$ and $v$ are covered by edges from $E\setminus \{e\}$, and there are exactly $x$ ways to cover all vertices in the gadget including $u$ and $v$. Furthermore, there is no cycle cover in the gadget that covers only one of $u$ and $v$. However, the emerging cycle covers contain many cycles of length $2$ and $1$. Therefore, for each prime, $p$, the immanants must be evaluated on permutations with cycle structures depending on $p$. So far it is not obvious that appropriate prime numbers $p$ can be selected such that the now varying
$$
\sum_{m=0}^k {k\choose m} \chi_{\lambda(p)}(\rho(p,k,m))
$$
is not divisible by $p$ for any $k$. Note that Valiant's gadget changes the number of vertices, thus the corresponding shape depends on $p$, and so the cycle structure of the relevant permutations on which the immanant must be evaluated.

\section*{Conclusion and outlook}
We have shown that a broad family of immanants $\operatorname{\Imm}_{\chi_\lambda}$ is $\#P$-hard to compute on two highly structured input classes: $0$-$1$ matrices and adjacency matrices of edge-weighted planar directed graphs. In the planar case the edge weights are nonnegative integers bounded by a quadratic function of the number of vertices.

The source of hardness is representation–theoretic, governed by the shape of $\lambda$; in particular, the presence of a sufficiently large domino--tileable region (under mild technical conditions) already forces $\#P$-hardness.

It is natural to place our results in the broader setting of \emph{class–function immanants}.
Given a class function $f:S_n\to\mathbb{C}$, the $f$-immanant of $X=(x_{ij})$ is
\[
\operatorname{Imm}_{f}(X)\;=\;\sum_{\sigma\in S_n} f(\sigma)\,\prod_{i=1}^n x_{i,\sigma(i)}\,,
\]
and, since the irreducible characters $\{\chi_\mu\}_{\mu\vdash n}$ form a basis of class functions, we may write
$f=\sum_{\mu\vdash n} c_\mu\,\chi_\mu$.
This framework subsumes the familiar cases: $\operatorname{per}(X)$ (when $f\equiv 1$), $\det(X)$ (when $f=\mathrm{sgn}$), and the character immanants $\operatorname{Imm}_{\chi_\mu}(X)$.
Motivated by our results it is natural to ask analogous complexity and approximability questions for other choices of $f$; we highlight the fermionant and the Temperley--Lieb immanants below.

\paragraph{Fermionants.}
A well-studied example is the \emph{fermionant}~\cite{Chandrasekharan2011},
\[
\operatorname{Ferm}_N(X)\;=\;\sum_{\sigma\in S_n}(-N)^{\mathrm{cyc}(\sigma)}\prod_{i=1}^n x_{i,\sigma(i)},
\]
where $\mathrm{cyc}(\sigma)$ is the number of cycles of $\sigma$. Note that $\operatorname{Ferm}_{1}(X)=(-1)^n\det(X)$ and $\operatorname{Ferm}_{-1}(X)=\mathrm{per}(X)$. Mertens and Moore~\cite{MM2013} proved that for any fixed integer $N>2$, computing $\operatorname{Ferm}_N(A)$ for the adjacency matrix $A$ of a planar graph is $\#P$-hard. They also observed, via the  known inapproximability of the Tutte polynomial on planar graphs~\cite{GoldbergJerrum2012}, that for any fixed $N>2$ the fermionant does not admit an FPRAS unless $\mathrm{NP}=\mathrm{RP}$. In contrast, there \emph{is} an FPRAS for the permanent on nonnegative matrices~\cite{jsv}. Moreover, by Valiant’s reduction~\cite{Valiant-permanent}, computing the permanent on entries from $\{-1,0,1,2,3\}$ is AP-reducible from $\#\mathrm{3SAT}$, so no FPRAS exists in that setting unless $\mathrm{RP}=\mathrm{NP}$. Beyond these cases, very little is known about the approximability of other immanants, even on restricted input classes.

\paragraph{Temperley--Lieb immanants.}
Temperley--Lieb (TL) immanants, introduced by Rhoades and Skandera~\cite{RhoadesSkandera2005}, are obtained by evaluating class functions $f_\tau:S_n\!\to\!\mathbb{R}$ arising from the Temperley--Lieb algebra (basis indexed by noncrossing matchings/diagrams) and forming $\Imm_{f_\tau}(X)=\sum_{\sigma} f_\tau(\sigma)\prod_i x_{i,\sigma(i)}$. They prove that these polynomials are \emph{totally nonnegative} (nonnegative on every totally nonnegative matrix) and that the cone they generate contains the classical $2{\times}2$ Plücker--type differences of products of minors $\Delta_{J,J'}(x)\Delta_{L,L'}(x)-\Delta_{I,I'}(x)\Delta_{K,K'}(x)$, along with new criteria on index sets guaranteeing total nonnegativity~\cite{RhoadesSkandera2005}. Their article~\cite{rhoades-skandera-fpsac04} gives an accessible summary in the planar-network/ Lindström --Gessel -- Viennot framework and situates TL immanants among previously known totally nonnegative $f$-immanants. To the best of our knowledge, no computational classification is known for TL immanants on $0$-$1$ inputs or on adjacency matrices of edge-weighted planar directed graphs with nonnegative weights bounded by $\mathrm{poly}(n)$; their approximability on nonnegative inputs is likewise open. Recent work of Lu, Ren, Shen, and Wang introduces a related family called \%-immanants and proves structural relations showing when a Temperley--Lieb immanant coincides with (or decomposes into) a short linear combination of \%-immanants via complementary–minor identities; potential algorithmic consequences for our restricted inputs remain open~\cite{LuRenShenWang2023}.
\paragraph{Partition–algebra generalization.}
A related generalization, due to Campbell~\cite{campbell}, introduces \emph{recombinants}: immanant-like invariants built from partition-algebra characters. They specialize to ordinary immanants on the symmetric-group boundary and are defined as sums over partition diagrams rather than permutations. Their computational status appears open: we are not aware of algorithms or hardness results for recombinants on $0$-$1$ inputs or on adjacency matrices of edge-weighted planar directed graphs with nonnegative, polynomially bounded weights. It is natural to ask whether the planarity-preserving, cycle-type–pinning techniques developed here extend to this setting.

\section*{Acknowledgements}
The work presented in this article was initiated during the first author’s stay in Tromsø, funded by the Pure Mathematics in Norway project. The authors gratefully acknowledge the Tromsø Research Foundation for supporting this visit. In addition, the first author was supported by the Hungarian NKFIH grants K132696 and SNN135643, while the second author was supported by the Tromsø Research Foundation grant 17MatCR. The authors also thank Radu Curticapean for valuable discussions on an earlier version of this paper and for helpful explanations of the ideas underlying his construction. Moreover, the contributions and insights provided by two anonymous referees are  appreciated.
\bibliographystyle{plain}

\end{document}